\documentclass[twocolumn]{aa} % for a paper on 2 column
\usepackage{graphicx}
\usepackage{natbib}
\usepackage{latexsym}
\usepackage{amsmath}
\usepackage{amssymb}
\usepackage{amstext}
\usepackage{color}
\usepackage{psfrag}
\def\cm3{cm$^{-3}$}

\def\3{$^{13}$CO}
\def\2{$^{12}$CO}

\begin{document}
\title{First high-resolution radio study of the Supernova Remnant G338.3-0.0 associated with the 
gamma-ray source HESS~J1640$-$465}

\author{G. Castelletti\inst{1,2} \footnote[1]{Member of the Carrera del Investigador Cient\'\i fico of CONICET, Argentina}
\and E. Giacani\inst{1,3} \footnotemark[1]
\and G. Dubner\inst{1} \footnotemark[1]
\and B. C. Joshi \inst{4}
\and A. Pramesh Rao \inst{4}
\and R. Terrier\inst{5}
}

\institute{Instituto de Astronom\'\i a y  F\'\i sica del Espacio
(CONICET-UBA), CC 67, Suc. 28, 1428
     Buenos Aires, Argentina\\
     \email{gcastell@iafe.uba.ar}
\and
Facultad de Ciencias Exactas y Naturales, Universidad de Buenos Aires, Argentina 
     \and
Facultad de Arquitectura y Urbanismo, Universidad de Buenos Aires,  Argentina 
\and
       National Centre for Radio Astrophysics,
       Ganeshkhind, Pune 411007, India
\and
APC-Univ. Denis Diderot Paris 7, 75265 Paris Cedex 13, France}

     \offprints{Castelletti G.}

      \date{Received <date>; Accepted <date>}

%\abstract{}{}{}{}{}
 % 5 {} token are mandatory
 \abstract
%Context optional
% {We have performed a multifrequency study of the SNR ...... Our prime motivation
%was to detect .... Our basic aim was to ....}
{}
%Aims
 {To perform a multifrequency radio study of the supernova remnant (SNR) G338.3$-$0.0,
in positional coincidence with the TeV source HESS~J1640$-$465. To study the morphological and 
spectral properties of this remnant and its surroundings searching for plausible radio counterparts 
to the gamma-ray emission.}
%Methods
{To carry out this research we observed the SNR~G338.3$-$0.0 using the Giant Metrewave Radio
Telescope (GMRT) at 235, 610, and 1280~MHz. We also reprocessed archival data from 
the Australia Telescope Compact Array (ATCA) at 1290 and 2300~MHz. 
Also we conducted a search for radio pulsations towards a central point-like source, using the GMRT 
antennas at 610 and 1280~MHz. 
The molecular material in the region of the SNR was investigated based on observations made with the
NANTEN telescope in the $^{12}$CO~(J=1-0) emission line.}
%Results
{The new radio observations revealed a remnant with a bilateral 
morphology, which at 235~MHz has the western wing completely attenuated because of absorption due to foreground ionized gas. 
The quality of the new images allows us to provide accurate estimates for the total radio flux density of the whole SNR at different radio frequencies. From the new and existing flux density estimates between 235 and 5000~MHz we derived for the whole remnant 
a spectral index $\alpha=-0.51\pm0.06$ with a local free-free continuum optical depth at 235~MHz $\tau_{235}=0.9\pm0.3$.
No radio pulsations were detected towards the only radio point-like source within the HESS error circle. We derived  upper limits of 2.0 and 1.0~mJy at 610 and 1280~MHz, respectively, for the pulsed flux towards this source. 
No radio counterpart was found for the pulsar wind nebula
discovered in X-rays. 
The inspection of the interstellar molecular gas towards G338.3$-$0.0 and surroundings revealed that there is not any associated dense cloud that might explain a hadronic origin for the TeV detection.}
%conclusions optional  
{}


\titlerunning{First radio study of SNR~G338.3$-$0.0}
\authorrunning{\textsc{Castelletti et al.}}

\maketitle
          %
          %________________________________________________________________
\section{Introduction}

The very high energy (VHE) $\gamma-$ray source \object{HESS~J1640$-$465} was discovered by the H.E.S.S. Cherenkov array during the Galactic survey in 2004$-$2006 \citep{aha06a}. With an rms width of 
2$^\prime.7 \pm 0^\prime.5$, it is among the most compact sources in that survey. The spectrum is hard, with a photon index $\Gamma \simeq2.42$, and a total integrated 
flux above 200~GeV of $\sim 2.2 \times 10^{-11}$~erg~cm$^{-2}$~s$^{-1}$ \citep{aha06a}. 
In the same area, \citet{sla10} reported the detection with the {\it Fermi}-Large Area Telescope of the source \object{1FGL~J1640.8$-$4634} likely arising from HESS~J1640$-$465.

Based on \it XMM-Newton \rm observations, \citet{fun07} identified a slightly extended 
(size $\sim$0$^{\prime}$.45) hard spectrum X-ray emitting
source (XMMU~J164045.4$-$463131) at the centroid of the HESS source.
This field has been also investigated using the \it Chandra \rm X-ray 
Observatory \citep{lem09} resolving the diffuse emission previously detected by \it XMM-Newton \rm and revealing a bright point source inside (centered at J2000 
R.A.=$16^\mathrm h40^\mathrm m43^\mathrm s.52$, dec=$-46^\circ31^\prime 35^{\prime\prime}.4$), labeled by the authors as S$_ {1}$. 
Based on the morphology and the X-ray spectrum the authors 
suggest that the X-ray point source is a putative pulsar and the diffuse component its 
associated pulsar wind nebula (PWN). \citet{sla10} interpret the {\it Fermi}-LAT observations 
as primarily originated in such a PWN. 

The VHE source lies in the interior, close to the borders of \object{G338.3$-$0.0}, a shell-like radio supernova remnant (SNR). 
The remnant, about 8$^\prime$ in size, is  located in a complex region of the sky, rich in both non-thermal and thermal emission \citep{cas87}. 
Based on HI absorption measurements and assuming that the SNR is at the same distance as 
the neighboring HII regions, a distance between $\sim$8.5 and 13~kpc was 
suggested for G338.3$-$0.0 \citep{lem09}. 
Following \citet{lem09} we will assume for our calculations a distance to 
G338.3$-$0.0 of 10~kpc, leaving our results expressed in terms of $d_{10}=d/10~\mathrm{kpc}$.

Up to now, the observations carried out failed to unambiguously
confirm the connection between G338.3$-$0.0 and HESS~J1640$-$465. 
In this paper we present new and reprocessed data at 235, 610, 1280, and 2300 MHz acquired with
the Giant Metrewave Radio Telescope (GMRT) and ATCA, with the aim of investigating the SNR and
searching for a radio counterpart for the proposed pulsar and PWN observed in X-rays.
 
\section{Observations and data reduction}

\subsection{GMRT observations}
We have performed low frequency observations of the SNR~G338.3$-$0.0 at 235, 610,
and 1280~MHz using the GMRT. Simultaneous dual frequency observations were carried out at 
235 and 610~MHz during two 6~hours sessions on 26 and 27 February 2008.    
Data in the 1280~MHz band was obtained on 29 February and 1 March 2008 for a total 7
hours observing time on G338.3$-$0.0.  
To make feasible radio frequency interference (RFI) excision on the data, the observations
at 610 and 1280~MHz were performed for both the upper and lower sidebands each of which 
with a total bandwidth of 16~MHz split into 128 spectral channels. At 235~MHz the data were 
collected only for the upper sideband
using 64 channels over a total bandwidth of 8~MHz.

\begin{table*}
\caption{Properties of the full-resolution images of SNR G338.3$-$0.0}
\renewcommand{\arraystretch}{1.0}                                                                                
\begin{center}
\begin{tabular}{cccc} \hline\hline 
Frequency &  Synthesized beam &  Beam position  &  Noise level\\
(MHz)     &   (arcsec)        &   angle (deg)  &  (mJy~beam$^{-1}$)\\ 
\hline
235     & 25.8 $\times$ 9.6 &      8.0     & 18       \\
610     & 12.6 $\times$ 5.0 &    $-$0.8    & 1.6    \\
1280    & 6.4  $\times$ 5.0 &      56.9    & 1.3    \\
2300    & 4.8  $\times$ 4.3 &     7.7     & 0.44  \\  \hline     
\end{tabular}
\end{center}
\label{table:beams}
\end{table*}

At the three frequencies, the absolute amplitude scale was set using the calibrator \object{3C~286}.
Bandpass calibration was in all cases achieved by observing the source \object{1830$-$360}.
In the 235/610 MHz observations, 1625$-$311 was used as a phase calibrator, 
whereas \object{1626$-$298} was the secondary calibrator for the 1280 MHz data.
Automatic measurements of the variations in the system temperatures of the antennas at
the GMRT were not implemented during the observations. Such changes in temperature may 
become significant when observing the Galactic plane region at low radio frequencies.
Additional correction factors (ranged from 1.1 to 2, with the higher value at 235 MHz)
applied to each dataset after imaging
were estimated  using interpolated values of sky temperatures from the 408~MHz all-sky continuum survey of \citet{has82} and assuming a spectral index $\beta=-2.7$ for the galactic background emission. 
 
The data from each day were fully reduced and imaged
separately to ensure that there were no day-to-day amplitude discrepancies.
In the case of the 610 and 1280 MHz observations, the data from each of
the two available sidebands were also reduced independently. 
To perform this work we made use of
the NRAO Astronomical Image Processing Software (AIPS) package following
a similar procedure for each observed frequency. 
After initial calibration the dataset from each sideband at 235, 610, and 1280~MHz were 
averaged in frequency by collapsing the bandwidth to a 
number of five, eleven, and three spectral channels, respectively. 
The individually calibrated 235 MHz datasets from each day were then concatenated.
At 610 and 1280~MHz, before combining data from each day,  the data recorded for the upper and
lower sidebands were combined into a single \it uv \rm dataset using the tasks UVFLP and BLOAT
within AIPS and imaged as outlined below \citep{gar07}. 

In order to reduce the \it w\rm-term effect imposed by the large field of view associated 
with low frequency radio observations (the primary beam of the telescope at 610 and 235~MHz 
is $\sim$0$^{\circ}$.8 and $\sim$2$^{\circ}$, respectively)
we employed wide-field imaging as implemented in the AIPS task IMAGR. 
To deal with ionospheric based phase variations, which decorrelate phases across the full field of view we employed several rounds of phase-only self-calibration and a final phase 
and amplitude self-calibration to the visibility data at 235, 610, and 1280~MHz. 
The final calibrated visibility data at 
235~MHz were combined into a single \it uv \rm dataset and imaged using a SDI Clean algorithm.
All the deconvolved images were finally corrected for primary beam attenuation.
The properties of the final images  are summarized in Table~\ref{table:beams}.

\subsection{ATCA archival data}
The 1280~MHz~GMRT observations  are sensitive to 
structures with angular scales up to $490\lambda$$\simeq7$~arcmin. To improve the \it uv \rm 
coverage we combined in the \it uv \rm plane the GMRT data with archival ATCA data acquired in the same radio band in November and
December 2005 with the array operating in two  different configurations (750D and 1.5C). 
The region was imaged in the mosaic mode with 5 pointing centers. The ATCA data were taken simultaneously at 1290 and 2300~MHz. For all observations, PKS~1934$-$638 and PKS~1740$-$517 were used as primary flux density and secondary phase calibrator, respectively. 
The shortest array spacing of the ATCA data was 31~m corresponding to a maximum angular 
scale to which the observations were sensitive as large as $\sim26$ and $\sim14$~arcmin at
1290 and 2300~MHz, respectively. 
The final angular resolution and sensitivity of
the image at 1280~MHz formed from the combination of GMRT and ATCA data and of the image at 2300~MHz
are listed in Table~\ref{table:beams}. 

Also, to investigate in detail the radio emission around the X-ray pulsar candidate in a search
for traces of a PWN, we  
reprocessed 4800 and 8640~MHz data corresponding to observations acquired with the ATCA telescope in the 750D configuration in 1999. 
These data are only useful for the study of structures smaller than 7~arcmin at 
4800~MHz and 4~arcmin at 8640~MHz, which correspond to the largest angular scale that can be
reasonably well imaged with the array at these high frequencies.

\section{Search for a radio pulsar}
From the archival ATCA observations carried out in the direction of G338.3$-$0.0 at 4800
and 8640~MHz, we detected a  point-like source located near the proposed X-ray
pulsar XMMUJ164045.4$-$463131 \citep{fun07}, unresolved down to the 
$3^{\prime\prime} \times 1^{\prime\prime}$ interferometric beam. This radio source, centered 
at R.A.=$16^{\mathrm{h}}\,40^{\mathrm{m}}\,48^{\mathrm{s}}$, 
dec.=$-46^{\circ}\,31^{\prime}\,58^{\prime\prime}$ (J2000), has a flux density of about 
6~mJy and 4.5~mJy at 4800 and 8640~MHz, respectively, and a radio spectral index $\alpha\sim-0.5$ ($S\propto\nu^{\alpha}$). 
It is worth mentioning that, at our current spatial resolution and sensitivity, this is the only radio point source present in the vicinity of the X-ray compact source. 
Though its radio spectral index is not typical for a pulsar, separate time series observations towards this point source were carried out nevertheless in a search for radio pulses for completeness.

The observations were conducted using the GMRT at 610 and 1280~MHz on 24 and 25 February 2008, respectively. On both the days we used 25 antennas of GMRT configured in an incoherent array  with a total bandwidth of 16~MHz. The FWHM for this configuration was 40$^{\prime}$ and 20$^{\prime}$ for 610 and 1280~MHz respectively, centered at the point source mentioned above. This beam also covers the source  S$_ {1}$  discovered by \citet{lem09} one year later.  We also used 16 antennas located in the compact central square array of GMRT in a  phased array mode synthesizing a more sensitive beam with FWHM of about 40$^{\prime\prime}$ and 20$^{\prime\prime}$ at 610 and 1280~MHz respectively, centered at the point source. The time series data were sampled every 256~microsec. The data were simultaneously acquired with incoherent array and phased array. This choice of observing configuration not only allowed us  more sensitive observations of the point source using the phased array\rm, but also to look for other related sources in the primary beam  with the incoherent array\rm, such as the mentioned source S$_ {1}$. Observations at 610~MHz were motivated by a search for sources, which may be related to SNR, with a steeper spectra than the point source.

The data were analyzed using the SIGPROC\footnote{More information available at
 http://sigproc.sourceforge.net} pulsar data analysis software on 128 processors high performance computer cluster currently in use at the National Centre for Radio Astrophysics. The data were dedispersed for 512 trial dispersion measures (DMs) with DMs ranging from 0 to about 465~pc~cm$^{-3}$ at 610~MHz. DMs were spaced more coarsely at 1280~MHz, due to the smaller dispersion smearing across each individual frequency channel. At this frequency the dedispersion search was made in 160 steps using DMs between 0 to 660~pc~cm$^{-3}$. We searched for periodicities in the dedispersed time series using a standard harmonics search and used a threshold signal-to-noise (S/N) of 8. Interference periodicities corresponding to power line
RFI as well as other known RFI were eliminated. After a careful examination of the profiles for the candidates produced by search processing, we did not find any characteristic periodic signal, as expected from a pulsar.

A calibrator source, 1830$-$360, was also observed 
using both the incoherent and phased array. 
This source was used to estimated the
rms noise in the time series. Upper limit on flux density of any pulsar in our field of view with S/N of 8 times rms were estimated as 2 and 1~mJy for 610 and 1280~MHz, respectively. Therefore, 
though our observations did not reveal pulsations from an intense source, they
do not rule out a pulsar fainter than the limits quoted above. In addition, the pulsed radiation may be beamed away from our line of sight. 

With a spectral index of $-$0.5, much less steeper than the average pulsar spectral index of about $-$1.8 \citep{mkk00}, our limits on pulsed radiation imply that the radio point source, shown in Fig.~\ref{radio-imagenes} must be unrelated. Source S$_ {1}$  may be a radio pulsar, but is either fainter than our limits or beamed away. Lastly, the high sky background and large interstellar scattering at a distance of 10~kpc could also lead to non-detection of pulsed radiation, particularly for periods below 100~ms, due to scatter-broadening of the radio pulse.

\section{New radio images of G338.3$-$0.0}
\begin{figure*}[ht*]
  \centering
\includegraphics{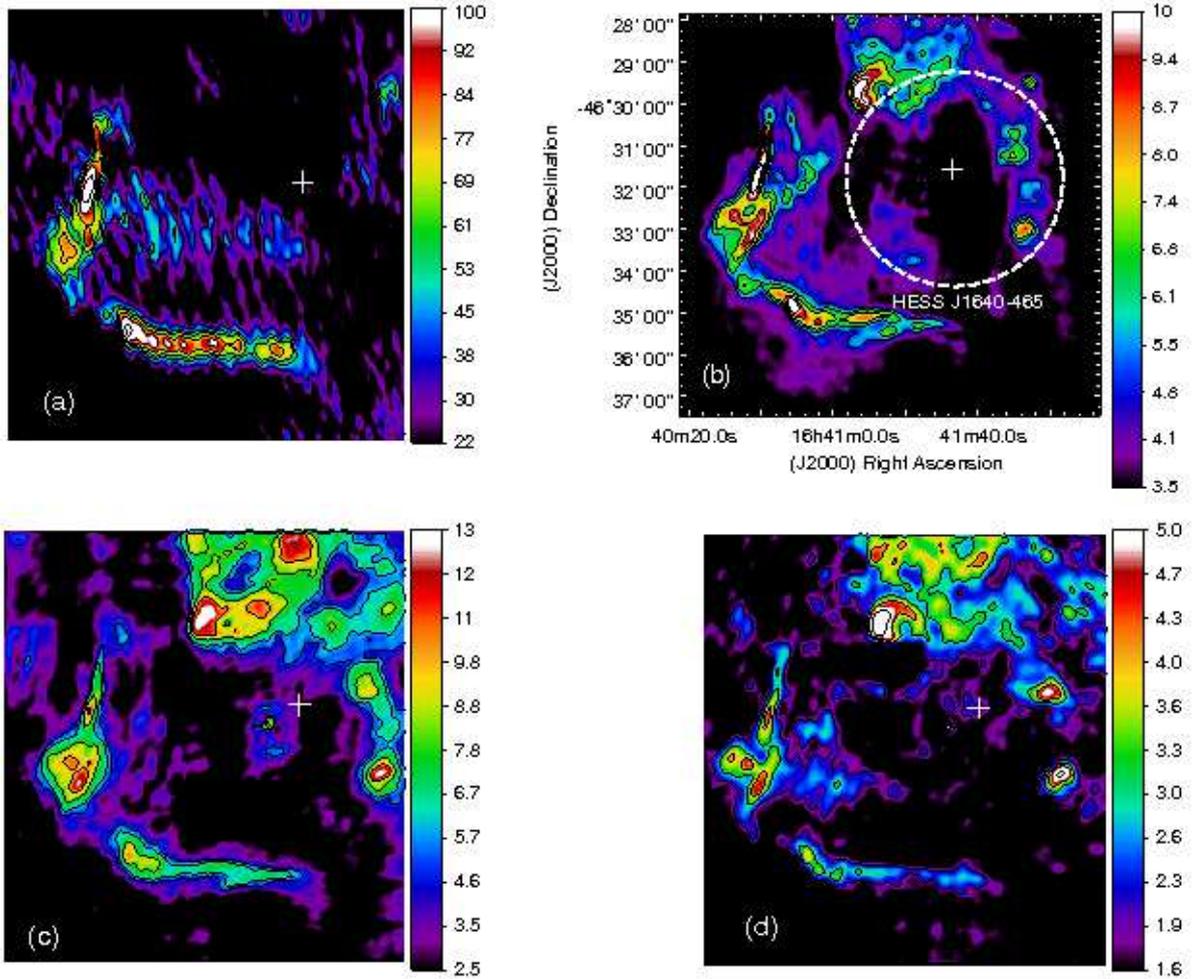}
 \caption{New radio continuum images from the SNR~G338.3$-$0.0.
The contours and the linear intensity scales have been chosen  
to emphasize the relevant features at each frequency. 
The wedges display in mJy~beam$^{-1}$ the values of the radio emission at each frequency. 
All images were aligned to the same pixel position. The coordinates displayed in Fig.~1b are the same for the other three panels. 
\bf a) \rm GMRT image at 235~MHz. 
The emission towards the western hemisphere is significantly
suppressed at this frequency. The contours are traced at 36, 50, 60, 74, 90, and 
120~mJy~beam$^{-1}$. 
\bf b) \rm GMRT image at 610~MHz. 
The intensity contours correspond to 3.5, 5, 6, 7, 8.5, and 10~mJy~beam$^{-1}$. 
\bf c) \rm Radio emission at 1280~MHz from GMRT and ATCA data. 
The contours are traced at 4, 5, 6, 8, and 10~mJy~beam$^{-1}$. 
\bf d) \rm 2300~MHz image obtained from reprocessed archival ATCA data.
The radio contours are traced at 2, 3, 4, and 6~mJy~beam$^{-1}$.
All the displayed images include primary beam correction. 
The synthesized beams, position angle and noise levels are listed in Table~\ref{table:beams}.
The white dashed circle included in Fig. 1 (b) shows the position and size of the VHE gamma-ray source  HESS~J1640$-$465.
The plus sign in each panel marks the position of the X-ray source S$_{1}$ identified by \citet{lem09}.
(A color version of this figure is available in the on line journal).}
\label{radio-imagenes}
\end{figure*}

In Fig.~\ref{radio-imagenes} we present close-up views of 
a portion of the field mapped with GMRT at the low radio frequencies plus the ATCA reprocessed data, showing the detailed morphology of G338.3$-$0.0 at 235, 610, 1280, and 2300~MHz. 
These are the 
first high-resolution images of the source providing unprecedented sensitivity to the radio emission. The fact that the smallest baseline in the GMRT Central Square for which the data at 235 and 610~MHz were reliable correspond to $\sim86\lambda$ and $\sim230\lambda$ respectively, 
equivalent to structures with angular scales up to 40 and 15~arcmin,
ensures that the contributions from the largest spatial scales have been
adequately recovered for this SNR about $8^{\prime}$ in size. 
Particularly, the GMRT~610~MHz image (Fig.~\ref{radio-imagenes}b) improves by a factor 6 the angular resolution and 7 times the sensitivity in comparison with the only previously known image of this remnant at 843~MHz \citep{whi96}.
In the case of the data at 1280~MHz, since G338.3$-$0.0 is almost at the limit of the largest imaged structures
we incorporated the shortest spacings information as taken from archival ATCA observations to avoid missing large scale flux density in the image at this frequency, as described in Sect.~2.2. 

Except for the radio image at 235~MHz, where most of the western half is completely 
attenuated, the total intensity images at 610, 1280, and 2300~MHz are quite similar in appearance, revealing a supernova remnant with a clear bilateral morphology. Considerable small scale features, including bright knots, are evident in the
new images along the entire shell, with their respective brightness varying across
the spectrum. Diffuse emission is also observed inside the radio shell, particularly close
 to the eastern radio shell. 

\begin{figure}[ht]
\centering
%\includegraphics{figures-g338/610-hii-subim-v3.epsi}
%\includegraphics[totalheight=0.3\textheight,viewport=0 0 360
%235,clip]{hii-snr-negro.eps}
%\includegraphics[width=3cm]{hii-snr-negro.ps}
%\includegraphics[width=8cm]{figures-g338/figure2-bb.epsi}
\includegraphics[width=8cm]{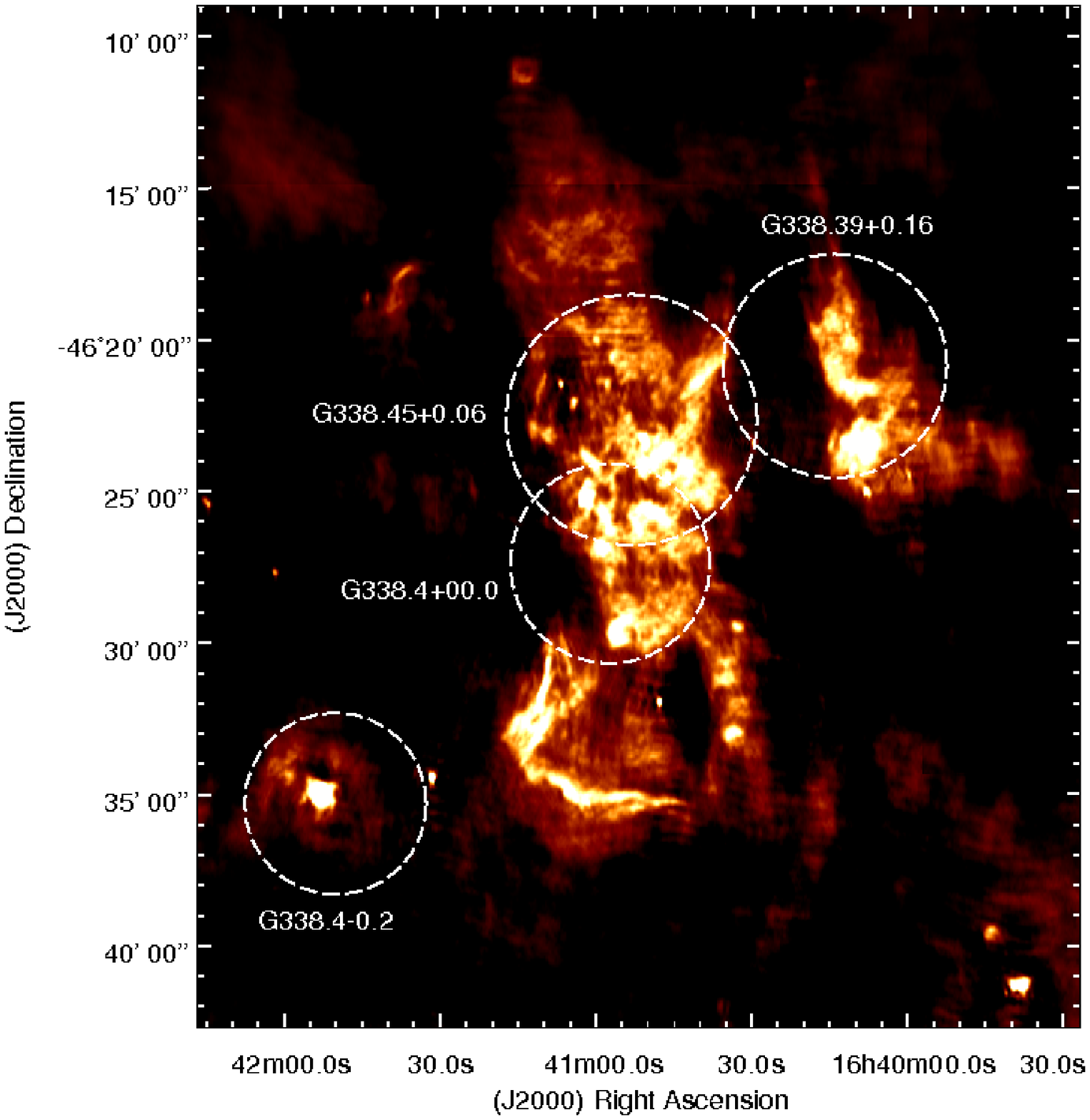}
\caption{Radio continuum image at 610~MHz of a large area around the SNR~G338.3$-$0.0, showing several nearby thermal sources encircled by dashed white lines to facilitate their location. 
The intensity scale is based on a linear relation from 2.4 to 8~mJy~beam$^{-1}$. 
(A color version of this figure is available in the on line journal).}
\label{HII}
\end{figure}

The large sky field mapped at 610~MHz allows us to get a picture of the different sources emitting around G338.3$-$0.0.
In Fig.~\ref{HII} we present the radio emission at 610~MHz in an area of about 0.5 square degrees.
Several HII regions are evident in the field. The position and extension of the cataloged thermal regions (from Caswell $\&$ Haynes 1987 and Paladini et al. 2003) are indicated in Fig.~\ref{HII} by white  dashed circles. From this figure it is obvious that the remnant
lies in a complex region in our Galaxy with an intricate network of non-thermal
and thermal structure, with ample regions of the SNR shell covered by the HII
regions G338.45+0.06 and G338.4+00.0. 

\begin{figure}[ht]
\centering
%\includegraphics[totalheight=0.3\textheight,viewport=0 0 360
%235,clip]{hii-snr-negro.eps}
%\includegraphics[width=7cm]{channel2-235-2-rgb.eps}
%\includegraphics[width=7cm]{figures-g338/613radio-cont-ir24-8.epsi}
%\includegraphics[width=7cm]{figures-g338/IR-8green-24red-radio610sum-blue-cmyk.labels.eps}
%\includegraphics[width=7cm]{figures-g338/IR-8green-24red-radio610sum-blue-rgb.labels.epsi}
%\includegraphics[width=7cm]{figures-g338/ir-radio-new-cmyk.eps}
\includegraphics[width=7cm]{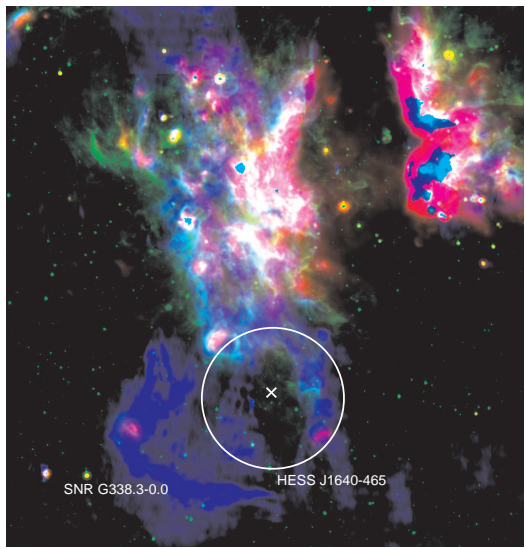}
\caption{A comparison of the radio continuum emission at 610~MHz (in blue) from the region
around SNR~G338.3$-$0.0 and the infrared emission at 8~$\mu$m (in green) and 24~$\mu$m (in red) 
taken from the \it Spitzer Space Telescope \rm GLIMPSE and MIPSGAL surveys, respectively.
The white circle marks the position and extension of HESS~J1640$-$465,
while the cross indicates the location of the X-rays source S$_{1}$ \citep{lem09}.
(A color version of this figure is available in the on line journal).}
\label{radio-ir}
\end{figure}

To accurately 
establish the flux density and hence the spectral properties of the SNR emission it is very important
to disentangle thermal from non-thermal contributions. To carry out this task we made use of 
the \it Spitzer \rm GLIMPSE and MIPSGAL infrared data.
In Fig.~\ref{radio-ir} we present a direct comparison of the IR emission 
with the new GMRT 610~MHz image in the spatial region around
G338.3$-$0.0. 
The emission in the 8 and 24~$\mu$m infrared bands are excellent tracers of warm dust associated
with star forming regions, usually related to thermal radio emission. 

Figure~\ref{radio-ir} clearly demonstrates that the entire complex of HII regions located to the north of G338.3$-$0.0  with an extent of about $12^{\prime}$ in the sky, is prominent in radio 
and in the two IR bands. In addition to the already noticed thermal emission, two infrared condensations particularly bright in 24~$\mu$m,  are observed overlapping the SNR. 
One located on the southern extreme of the western shell and the other near the center of the eastern 
half of the remnant. These sources with both strong infrared and radio emission are probably 
normal or ultracompact HII regions overlapping the non-thermal SNR emission. 
Other noticeable structure is a bright ring of infrared 8~$\mu$m and 24~$\mu$m emission located
on the northwestern portion of the SNR, which appears to outline precisely an enhancement
in the radio band. Such a feature was first identified at infrared wavelengths by
\citet{chu06} using GLIMPSE data who proposed that it is the projection of a fully
enclosed three-dimensional bubble (designated as S33 in Churchwell et al.'s catalog) with an 
angular diameter of about 0.9 arcmin.
The composite image also reveals the presence of faint infrared emission at 8~$\mu$m along the
northwestern boundary, within the error circle of the VHE source HESS~J1640$-$465.
We also investigated the near infrared emission using 
the 3.6~$\mu$m, 4.5~$\mu$m, and 5.8~$\mu$m 
Infrared Array Camera (IRAC) data from the GLIMPSE Science Program
but no emission above noise is detected in this region. 

With this information we attempt to provide accurate estimates for the flux densities over the whole 
SNR~G338.3$-$0.0 at 235, 610, 1280, and 2300~MHz. In addition we used the image of this SNR
acquired at 843~MHz with the MOST telescope
and the image at 5000~MHz taken from Parkes 64-m telescope archival data to extract another
values for the flux density.
Our estimate of the flux density at 843~MHz is in good agreement,
within uncertainties, with that reported by \citet{whi96}. 

These flux density estimates are corrected by a 
background level that was determined at each frequency
by tracing one-dimensional plots of the intensity as a function of position
in selected slices around the shell of the SNR.
We note a dependence on the direction of measurement of the background emission specially for
the images at 1280 and 2300~MHz, for which variations 
of a few tens of mJy~beam$^{-1}$ were calculated between
the east and west directions of the radio shell.
The values estimated in this way were then subtracted from the 
integrated flux density at each frequency.
An extra contribution from the thermal emission was also subtracted 
from the integrated flux density measurements. This contribution was estimated based on the
comparison of our radio image with the infrared data by
integrating the infrared flux in the bright regions overlapping the SNR and determining their
respective fluxes at the observed radio frequencies by 
assuming a spectral index of about
$\alpha\sim-0.1$ for the thermal emission.
Table~\ref{table:obs_radio} gives the integrated flux density measurements for G338.3$-$0.0 from 235 to 5000~MHz. 
In the second column we list the results obtained by directly measuring the flux
density from the image at each frequency. 
The listed flux densities represent the average from different measurements carried out using slightly
different contours to enclose the SNR emitting region.
In columns 3 and 4 we summarize the background and thermal emission contribution, while
the final flux density values after subtracting these contributions are listed in the 
fifth column of Table~\ref{table:obs_radio}. 
The results in this table have been placed on the same absolute flux density scale 
of \citet{baa77}. 

\begin{table*}
\caption{Flux density estimates for the SNR~G338.3$-$0.0}
\renewcommand{\arraystretch}{1.0}     
\begin{center}
\begin{tabular}{ccccc}\hline\hline
Frequency   & Directly measured & Background & Thermal & Final  \\ 
(MHz)       & flux density (Jy) & emission (Jy) &  emission (Jy) & flux density (Jy)\\ \hline
235 & $7.0 \pm 1.2$ & $0.5 \pm 0.2$ & $0.3 \pm 0.2$ & $6.2 \pm 1.2$ \\ 
610 & $8.80 \pm 0.94$ & $0.20 \pm 0.05 $ & $0.45 \pm 0.10$ & $8.15 \pm 0.95$ \\ 
843 & $8.6 \pm 0.7$ & $0.3 \pm 0.1$ & $1.2 \pm 0.4$ & $7.1 \pm 0.8$ \\ 
1280 & $8.5 \pm 0.7$ & $0.9 \pm 0.3$ & $1.3 \pm 0.2$ &  $6.3 \pm 0.8$ \\ 
2300 & $7.1 \pm 0.7$ & $1.1 \pm 0.4$ & $1.2\pm0.1$ & $4.8 \pm 0.8$ \\ 
5000 &$4.7 \pm 0.2$ & $0.3\pm 0.1$ & $1.3 \pm 0.1$ & $3.1 \pm 0.2$ \\ \hline
\end{tabular}
\end{center}
\label{table:obs_radio}
\end{table*}

\begin{figure}[h]
  \centering
\includegraphics [width=9cm]{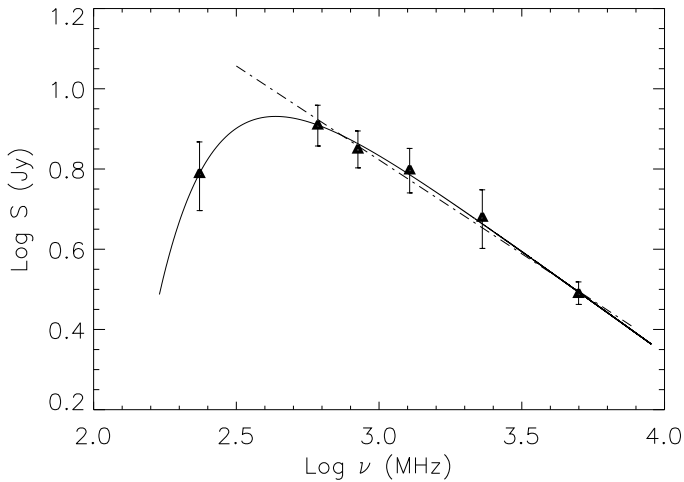}
        \caption{Integrated radio continuum spectra of the SNR~G338.3$-$0.0 obtained from the
flux density values listed in Table~\ref{table:obs_radio} (column 5). 
The dashed line is the best fit to the flux density measurements at frequencies higher than 235~MHz,
which gives an integrated spectral index $\alpha=-0.47 \pm 0.05$.
The solid line is the fit to the data using Eq.(\ref{flux}), which takes into account 
a low frequency turnover due to absorption by foreground thermal gas. This method yields a
global spectral index for G338.3$-$0.0 $\alpha=-0.51 \pm 0.06$.
}
\label{spectra}
\end{figure}

Figure~\ref{spectra} shows the radio continuum spectrum of G338.3$-$0.0 
from the flux densities summarized in Table~\ref{table:obs_radio}. 
There is a clear sign of a low frequency turnover in this spectrum 
likely to be caused by the emission associated with the intervening thermal gas. 
We first used a single power law spectrum excluding the flux density 
value at 235~MHz to determine the integrated spectral index of the remnant 
(represented as a dashed line in Fig.~\ref{spectra}). 
A weighted fit yields a spectral index $\alpha=-0.47\pm 0.05$ ($S\propto \nu^{\alpha}$), 
typical for shell-type SNRs. 
In addition, 
to determine the properties of the gas whose emission is 
responsible for the observed turnover,  
we have used all the integrated flux measurements, to make a weighted fit to the equation (indicated by the solid line in Fig.~\ref{spectra}) 

\begin{equation}
S_{\nu}=S_{408}\,\left(\frac{\nu}{408~\mathrm{MHz}}\right)^{\alpha}\,
\mathrm{exp}\left[-\tau_{408}\left(\frac{\nu}{408~\mathrm{MHz}}\right)^{-2.1}\right] 
\label{flux}
\end{equation}

\noindent
where $S_{\nu}$(Jy) represents the flux density at the frequency $\nu$(MHz). $S_{\mathrm{408}}$ and
$\tau_{\mathrm{408}}$ are the flux density and the average optical depth at the fiducial frequency of 408~MHz.
This equation assumes a constant spectral index $\alpha$ for the synchrotron emission over the entire 
radio band, allowing for a turnover at the lowest frequencies \citep{kas89}.
The best fit for the parameters are  $S_{\mathrm{408}}=11.2\pm1.3$~Jy 
and $\alpha=-0.51\pm0.06$, whereas
the best fit value for the free-free continuum optical depth is $\tau_{\mathrm{408}}=0.28\pm0.08$.
The average optical depth at 235~MHz estimated
from $\tau_{\nu}=\tau_{408}\,[\nu/408]^{-2.1}$ is $0.9\pm0.3$, which should be considered as
an upper limit due to the fact that the low frequency turnover observed in
the spectrum of Fig.~\ref{spectra} occurs near the lowest observed frequency,
resulting in a possible  overestimation \citep{kas89}. 
Accurate flux density measurements of G338.3$-$0.0 at very low
frequencies ($\leq100$~MHz) are highly desirable to better establish the frequency dependence of
the spectrum in this region.
The derived spectral properties of G338.3$-$0.0 are not surprising given that, 
as we have shown in Fig.~\ref{HII}, the remnant lies in a site that hosts numerous HII regions. 
Thus, contamination from ionized gas from these HII regions (or their envelopes) along the line of sight constitutes a potential candidate to provide extrinsic free-free absorption towards the SNR.

By applying appropriate \it uv-\rm tapering, in order to match the range of spatial scales 
measured at 235 and 610~MHz, we studied the dependence on
the position across the face of the remnant of the optical depth at the lowest frequency. As expected the higher
values of this magnitude correspond to the western and northwestern part of the SNR where
$\tau_{235}$ varies between 1 and 2. Based on this result together with our spectral analysis,
we can derive the physical properties 
of the thermal plasma responsible for the observed attenuation in the western part of the SNR.
We conclude that if the electron temperature varies in the typical range
for HII regions (5000 to 10000~K) and if the absorbing gas has a deepness along the line of sight
comparable to its transverse size, at the assumed distance of 10~kpc, the emission measure varies 
between 0.1 and 0.3$\times10^{6}$~pc~cm$^{-6}$ in the region where the absorption is stronger,
and the electron density is in the range 100 to 165~cm$^{-3}$.

\section{Search for a PWN in G338.3$-$0.0}
Based on the new radio images 
we have 
searched for a counterpart to the observed X-ray emission interior to the SNR radio shell, proposed
to be  a PWN \citep{fun07,lem09}.
However, down to the noise level of our data at each frequency  they do 
not reveal any trace of radio emission associated with the X-ray PWN. On the contrary, instead of the expected nebular enhancement, a depression in the emission is observed at 610~MHz in the area covered by the X-ray nebula.
In spite of the complex nature of the emission in this region, there is no reason to suspect a deficit
in the recovered flux density, since observations at this frequency were made with very good spatial
coverage and no overlap of thermal emission is evident in coincidence with the X-ray nebula. 
We can therefore only set limits on the radio flux density assuming for our 
calculation that particles originate close to the X-ray point source 
and then disperse in a nebula whose extent in radio is, at least, comparable to that 
of the X-ray nebula in G338.3$-$0.0. 
The best constraint is $\sim6$~mJy obtained from the image at 610~MHz as it represents the lowest between all the flux density measurements. 

Assuming a PWN radio spectral index of $\alpha=-0.3$ \citep{gae00}, the mentioned upper limit corresponds to a broadband radio luminosity integrated between $10^{8}$ and $10^{12}$~Hz of 
$L_{\mathrm{R}}\sim6\times10^{32}\,d_{10}^{2}$~erg~s$^{-1}$, where the upper cutoff frequency 
considered in this calculation is obtained from the spectral break between radio and X-rays from
the \it XMM-Newton \rm observations \citep{fun07}. 
In this case, the derived radio luminosity for this object is distinctly lower than
$L_{R}\sim10^{34}$~erg~s$^{-1}$, observed in other radio PWNe \citep{gae06}. On the contrary, a higher luminosity of $L_{R}\sim9.0\times10^{34}\,d_{10}^{2}$~erg~s$^{-1}$ is obtained if, the upper cutoff is derived from the fit between radio and \it Chandra \rm data
\citep{lem09}, i.e. a lower and upper cutoff frequency of $10^{8}$  and  $10^{16}$~Hz
are used in the calculation. This latter result, however, is at odds with the non-detection of 
a radio nebula in the region.
The dramatic difference between the calculated luminosities comes from the uncertainty in the frequency at which the spectral
break between radio and X-rays takes place, which is due to the quite different X-ray slopes quoted in the literature.
In order to reproduce the broad-band emission 
\citet{fun07} proposed a time-dependent rate injection of relativistic electrons in which the 
X-ray emitting region is originated by youngest particles ($t<2000$~yr), while an older population of electrons injected early in the formation of the PWN dominates the
VHE $\gamma$-ray emission. 
In this scenario if the population of youngest electrons also dominates the radio emission, the values of the radio flux densities would fall well below the sensitivity limit of current observations,  compatible with our non-detection of the radio PWN.
 
\section{The environmental molecular gas}
In a search for an alternative explanation for the observed $\gamma$-ray emission, 
we analyzed the $^{12}$CO~(J=1$-$0) molecular line emission around G338.3$-$0.0 based on data 
taken from the CO survey carried out with the 4-m NANTEN 
telescope operated in Las Campanas Observatory (Chile). 
These observations 
have a HPBW of 2.7~arcmin, a grid spacing of 4~arcmin, and a velocity resolution of 1~km~s$^{-1}$. 
 
After carefully inspecting the complete $^{12}$CO data cube 
we did not find any molecular cloud that, based on morphological correspondence,
might be associated with either the source HESS~J1640$-$465 or the SNR radio shell, 
at least down to the sensitivity and resolution of the
CO survey used. Deeper molecular observations would be very useful to investigate the
likelihood of a hadronic origin for the observed $\gamma$-ray emission.

\section{Summary}
This paper presents the highest angular resolution and sensitivity multifrequency dataset of the radio
continuum emission towards the SNR~G338.3$-$0.0. A mixture of synchrotron radiation from the remnant and overlapping unrelated thermal gas 
it is noticeable in the mapped region.
The quality of the new images provides for the first time proper
estimates for the total radio flux density of the whole SNR at four radio frequencies. 
Based on these images we studied the global radio continuum spectrum of the 
remnant and found a clear indication of a low frequency turnover. 
We conclude that a foreground sheet of ionized gas is absorbing the radio synchrotron emission coming from almost half of the radio SNR shell. We interpret that this absorption is associated with the complex of HII regions in close proximity to the remnant.

We also identified a previously  uncataloged point-like radio source near the centroid of the TeV source
HESS~J1640$-$465, close to the compact object  detected in X-rays \citep{lem09} and considered by these authors as a putative pulsar which would be powering a PWN detected in X-rays.
No pulsed emission associated with the radio point source, or within the field of view, was detected up to a continuum flux density of 2.0 and 1.0~mJy at 610 and 1280~MHz, respectively. 
Besides, no nebular radio emission
was detected in correspondence with the X-ray PWN in spite of the good quality of our 
radio images down to low surface brightness limits. In this context, we only set an upper limit on the radio emission at 610~MHz over a region with an area comparable to that of the PWN seen in X-rays. 

Finally, we analyzed the environmental conditions for a hadronic origin of the $\gamma$-ray emission. 
We concluded  that no bright molecular matter is apparent overlapping either the source
HESS~J1640$-$465  or the SNR radio shell.

In summary, these  radio observations fail to provide further support to the 
association of the SNR and/or its PWN with the emission observed in the $\gamma$-rays range. 
Physical mechanism producing only high energy photons should be proposed to reconcile the observations across the electromagnetic spectrum.

\acknowledgements{We acknowledge the referee for his/her constructive comments.
We thank Prof. Fukui Y. for kindly providing us with the NANTEN data. GMRT is operated by the National Centre for Radio Astrophysics of the Tata Institute of 
Fundamental Research. Pulsar search processing was carried out on NCRA's High Performance
Cluster and authors acknowledge assistance from V. Venkatsubramani and his cluster team. This work is based, in part, on observations made with the Spitzer Space Telescope, which is operated by the Jet Propulsion Laboratory, California Institute of Technology under a contract with NASA.
G.G. acknowledge financial support from User Community Development. 
This research has made use of the NASA's ADS Bibliographic Services.
Data processing was carried out using the HOPE~PC cluster at IAFE.
This research was partially funded through the following grants: CONICET (Argentina) PIP~112-200801-02166,
ANPCYT-PICT (Argentina) 0902/07, ANPCYT-PICT (Argentina) 2008-0795, and UBACYT~A023.}

\bibliographystyle{aa}
   %\bibliographystyle{klunamed}
   % format of references provided by the review (.bst)
\bibliography{paper}
   % file containing the bibtex references (.bib)
\IfFileExists{\jobname.bbl}{}
{\typeout{}
\typeout{****************************************************}
\typeout{****************************************************}
\typeout{** Please run "bibtex \jobname" to optain}
\typeout{** the bibliography and then re-run LaTeX}
\typeout{** twice to fix the references!}
\typeout{****************************************************}
\typeout{****************************************************}
\typeout{}
}

%BCJ change
%Add the following reference
%\bibitem[\protect{Maron et al.} }{2000}]{mkk+00}
% Maron O., Kijak J., Kramer M.\& Wielebinski R, 2000, A\&AS, 147, 195
%BCJ change

\end{document}